\documentstyle[12pt,tighten,aps,eqsecnum]{revtex}      
\begin{document}
\draft
\title{Dual Symmetry in Gauge Theories}
\author{ A.~L.~Koshkarov\thanks{Electronic address:
koshkar@mainpgu.karelia.ru}}
\address{Physics Dept., University of Petrozavodsk\\
Retrozavodsk, 185640, Russia}
\maketitle
\begin{abstract}
Continuous dual symmetry in electrodynamics, Yang--Mills theory and 
gravitation is
investigated. Dual invariant which leads to  badly nonlinear motion
equations is chosen as a Lagrangian of the pure classical dual nonlinear
electrodynamics. In a natural manner some dual angle which is determined by
the electromagnetic strengths at the point of the time-space appears in the
model. Motion equations may well be   interpreted as the equations
of the standard Maxwell theory with source.  Alternative interpretation is
the quasi-Maxwell linear theory with magnetic charge.
Analogous approach is possible in the Yang--Mills theory. In
this case
the dual-invariant non-Abelian theory motion equations possess the same
instanton solutions as the conventional Yang--Mills equations have.
An Abelian two-parameter dual group is found to exist  in Gravitation.
Irreducible representations have been obtained: the curvature tensor was
expanded into the sum of twice anti-self-dual and self-dual parts.
Gravitational instantons are defined as (real) solutions to the usual
duality equations. Central-symmetry solutions to these equations are
obtained. The twice anti-self-dual part of the curvature tensor
may be used for
introduction of new gravitational equations generalizing  Einstein's
equations. However, the theory obtained reduces to the conformal-flat
Nordstr$\ddot {\mbox{o}}$m theory.
\end{abstract}
\section{Introduction}
The position of Gauge Theories in the modern physics is exceptionally
important. The very name  includes the adjective pointing out  to the
important quality
in the theories -- gauge symmetry. Such theories involve also other
symmetries, not less important ones. Dual symmetry has taken
its own place in the number as well. It does not seem to be honorable
enough.

One cannot say that the dual symmetry was studied little --
besides a lot
of articles there are also a few  books. Especially this topic has been
discussing very actively  when the magnetic monopole problem
claimed
attention of physicists. Another related topic is instantons and monopoles
in the non-Abelian theories.

Still, it should be once more emphasized  that this field has not been
studied
enough. The Dual Symmetry (DUSYa) is the stepdaughter of Field Theory
while Gauge Symmetry is the favorite one.

It will be reminded  -- the dual symmetry like gauge one is
a necessary attribute of gauge theories. And in this fact an idea of
Unification, passing through the modern physics, is displayed.
However, what could be said of duality in Gravitation? Nearly nothing! But
this is one of the important gauge theories. In this paper  it is shown
  that
there exists two-parameter dual group in Gravitation.

A lot of attempts to solve the magnetic monopole problem in electrodynamics
by means of duality
have been mentioned
to be undertaken. There are many references to this subject in~\cite{str}
and~\cite{mpol}.
More modern approaches can be seen,
 e.g., in~\cite{olmon,mon}.

Dual invariance of the pure electrodynamics equations was found rather long
ago.  Having at first appeared as a discrete symmetry, this invariance
was then described as continuous one (dual turning)
by G. Rainich~\cite{rai},
C. Misner and J. Wheeler~\cite{mw}.

The interesting and strange fact is that the pure Maxwell electrodynamics
equations are invariant under the dual turning, and the Lagrangian does
not possess such a symmetry.

And what about  introducing the dual symmetry in theory from the very
beginning in a standard manner, having chosen the dual invariant as a
Lagrangian? The obvious objection to this idea is that the motion equations
 are badly nonlinear. Nevertheless, it will be seen further that the
inserted dual symmetry models are of interest.
For example, there may be an opportunity to interpret
the nonlinear dual theory as a linear Maxwell
theory  with source. In this case  the original theory nonlinearity is
hidden into the source. Non-Abelian fields  can be considered according
to a similar scheme.

On the other hand, the dual-covariant electrodynamics equations
may well be
interpreted
as linear equations of the electrodynamics with magnetic charge. Transition
from the picture with electrical charge to the picture with magnetic one is
provided by means of local dual transformation. Local dual angle arises in
the theory  in a
natural  way. A similar angle (phase, complexion) was
discussed by C. Misner  and J. Wheeler~\cite{mw}.

Establishing the dual group in Gravitation enables us to speak about
gravitational instantons as solutions to the duality equations like it
happens in the Yang--Mills theory. Moreover, the instanton sector
is here far
richer than in the Yang--Mills case since there are two irreducible
representations of the dual group. In addition this fact
gives new
opportunities to obtain new gravitational equations, to construct the
Lagrangians.

\section{Dual Symmetry in Electrodynamics}
\subsection{The Dual Transformations}
In the modern form the dual transformations were introduced by
Rainich~\cite{rai}, and Misner and Wheeler~\cite{mw}:
$$
F'_{\mu \nu }=F_{\mu \nu }\cos\theta +*F_{\mu \nu }\sin\theta
\equiv e^{*\theta }F_{\mu \nu },
$$
\begin{equation}
\label{dtu}
*F'_{\mu \nu }=-F_{\mu \nu }\sin\theta +*F_{\mu \nu }\cos\theta
\equiv e^{*\theta }*F_{\mu \nu },
\end{equation}
where $*F_{\mu \nu }=1/2\varepsilon _{\mu \nu \rho \sigma }F^{\rho \sigma
}$.

In view of formal similarity of the dual conjugation operator and imaginary
unit and because of the fact that the transition operator from ${\bf F}$
to ${\bf F'}$
looks like turning in some plane as well,  it becomes evident
that
transformation (\ref{dtu})
is referred to as the dual turning.

The pure Maxwell electrodynamics equations with the conventional quadratic
Lagrangian
$$
{\cal L} =gF^{\mu \nu }F_{\mu \nu }\equiv g{\bf F}^2
$$
prove to be invariant under the transformations (\ref{dtu})
while the Lagrangian itself does not possess the invariance!
That is, there is
no symmetry but this  points out that the symmetry might be realized
nonlinearly
(compare with spontaneous symmetry breaking).
Well-known as though,
this rather unusual fact
did not claim due attention of the theorists.

Meanwhile, the authors of the book~\cite{str}
playing on invariance of equations but ignoring the Lagrangian
noninvariance, try to  introduce  the "conservative" dual  current
what seems to us
 to be not quite correct.

It would be quite in  spirit of the modern approaches to consider the
Lagrangian with the explicit dual symmetry. It is known there is the only
dual invariant which is also Lorentz one. It has not been considered
(as far as I know) as a Lagrangian what is quite clear -- it would lead to
the badly nonlinear motion equation. Nevertheless it is of interest to
consider  the model of pure electrodynamics with dual-invariant Lagrangian,
first, in virtue of the unusual and very nice properties of the model and,
second, since this essentially nonlinear theory may well relate in some
way both
 to the Maxwell electrodynamics   and to the magnetic charge problem.

One can
obtain the only dual-invariant expression  as follows.
Let us consider a complex quadratic form
$$
{\bf F}^2+i*{\bf FF}.
$$
This form is usually regarded in complexifying
${\bf F}$-space. It is invariant under the complex orthogonal
transformations $O(3,C)$~\cite{dnf}.
With respect to the dual group (\ref{dtu})
the form transforms as follows:
$$
{\bf F}^2+i*{\bf FF} =({\bf F'}^2+i*{\bf F'F'})e^{2i\theta }.
$$
First, this formula  establishes isomorphism between the dual rotations
and the ordinary phase transformations. Second, it is getting clear that
the transformations (\ref{dtu})
are not a subgroup of $O(3,
C)$.
That is, the complex orthogonal transformations should not mix up
with the dual ones. Now it is quite clear that
$$
({\bf F}^2)^2+(*{\bf FF})^2=inv.
$$
It is the only expression to be invariant under either the Lorentz or the
dual transformations. Occasionally, it should be noted that for any
component $F_{\mu \nu }=f$
$$
f+i*f=(f'+i*f')e^{i\theta }\quad\longrightarrow
f^2+*f^2=inv.
$$

Let the Lagrangian of the pure dual  electrodynamics be chosen as follows:
$$
      {\cal L}=\sqrt{({\bf F}^2)^2+(*{\bf FF})^2}.
$$
Variational principle brings about the equations
$$
\left(F^{\mu\nu}\frac{{\bf F}^2}{\sqrt{({\bf F}^2)^2+(*{\bf F}{\bf F})^2}} +
*F^{\mu\nu}\frac{*{\bf F}{\bf F}}{\sqrt{({\bf F}^2)^2+(*{\bf F}{\bf F})^2}}
\right),_\nu=0.
$$
It is seen that the local angle $\varphi$
determined by electromagnetic strength at the point of space-time appears:
$$
\cos \varphi=\frac{{\bf F}^2}{\sqrt{({\bf F}^2)^2+(*{\bf F}{\bf
F})^2}},\quad \sin \varphi=\frac{*{\bf F}{\bf F}}{\sqrt{({\bf
F}^2)^2+(*{\bf F}{\bf F})^2}}.
$$
And now the motion equations can be written in  a shorter form
 $$
 (F^{\mu\nu}\cos \varphi+*F^{\mu\nu}\sin\varphi),_\nu=0.
 $$
Denoting
 $$
 F^{\mu\nu}\cos \varphi+*F^{\mu\nu}\sin\varphi=e^{*\varphi(x)}F^{\mu\nu},
 $$
one can rewrite these equations in a more elegant way:
\begin{equation}\label{me}
(e^{*\varphi(x)}F^{\mu\nu}),_\nu=0.
\end{equation}
Adding the identity
\begin{equation}\label{me'}
 *F^{\mu\nu}{}{}_{,\nu }\equiv 0,
\end{equation}
we obtain the complete system (set) of dual electrodynamics equations
(\ref{me}),(\ref{me'})
which is essentially non-linear.

\subsection{Transformation Properties of the Model under the Dual Group}
Further it is written down how some values transform.
\begin{equation}\label{tr1}
F_{\mu \nu }=e^{-*\theta }F'_{\mu \nu },\quad
*F_{\mu \nu }=e^{-*\theta }*F'_{\mu \nu },
\end{equation}
$$
\cos\varphi=\cos(\varphi '+2\theta ),\quad\sin\varphi=\sin(\varphi '+2\theta),
$$
\begin{equation}\label{tr3}
e^{*\varphi}F_{\mu\nu}=e^{*\theta}(e^{*\varphi'}F'_{}\mu\nu).
\end{equation}
All these formulae could be obtained directly. We can see that the
left-hand sides of the system (\ref{me}),(\ref{me'}) transform in various
ways. The left-hand side of (\ref{me'}) transforms like (\ref{tr1}), i.e.,
{\it co-gradiently} with respect to the field $F_{\mu \nu }$.
As seen from (\ref{tr3}),
the left-hand side of (\ref{me})
 transforms
by means of invert transformation, i. e.  {\em contra-gradiently} with
respect to $F_{\mu \nu }$.

\subsection{ Dual Electrodynamics as a Maxwell System with Source}
One can rewrite the set (\ref{me}),(\ref{me'})
 otherwise, just resolving the first of the equations with
respect to $F^{\mu \nu }{}{}_{,\nu }$:
\begin{equation}\label{ist}
F^{\mu\nu},_\nu=\varphi,_\nu(F^{\mu\nu}\tan \varphi-*F^{\mu\nu})\equiv j^\mu,
\end{equation}
$$
*F^{\mu \nu }{}{}_{,\nu }\equiv0.
$$
One can consider such representation of (\ref{me}),(\ref{me'})
as a way to break down the symmetry since both sides
of (\ref{ist}) transform in     different ways   .  Thereby, the
equation (\ref{ist})
takes  the form of the Maxwell equation with source. Let us suppose
that
the nonlinear set (\ref{ist}) has a solution of the form
$$
F_{0\mu }={\bf E}\sim
\frac{1}{r},\qquad j^0=o\left(\frac{1}{r}\right),\quad r\to \infty.
$$
Then perhaps the
notation of the right-hand side of (\ref{ist}) as a current could be
justified.  It will be noted that genuine central-symmetry field makes the
 right-hand side of (\ref{ist}) to be equal zero. That is all right. But
it would be better to find  an {\em asymptotically} central-symmetric
solution.

Going on to speculate  on this matter, it could be noted that
quantization of
charge which is effective in such a theory would arise as a result of
quantizing the  nonlinear field theory.

\subsection{Instantons in Electrodynamics}
Instantons in electrodynamics are known to be absent because of the gauge
group topology triviality. Although the term "instanton" should be made
more precise.

Let us notice the motion equations are satisfied if the condition
\begin{equation}\label{inst}
F_{\mu\nu}\cos \varphi+*F_{\mu\nu}\sin\varphi=0
\end{equation}
is fulfilled. Let it be called the generalized instanton equation. Why
instanton it will be seen later when discussing an analogous equation in
the non-Abelian theory. It is convenient to denote
$
a={\bf F}^2,\,b=*{\bf FF}.
$
Then, projecting (\ref{inst})
onto ${\bf F}$,
one obtains $a^2+b^2=0$,
or, on account of the reality of the fields $*{\bf F}$ and ${\bf F}$
$a=0,\,b=0$.  Thus, the electrodynamical instantons are, e.g., plane waves.

\subsection{The Local Dual Transformations}
Appearance of the angle $\varphi(x) $
in a natural manner prompts to introduce the space-time point-dependent dual
transformations. So
\begin{equation}\label{locdu}
\tilde F^{\mu\nu}=e^{*\varphi}F^{\mu\nu}.
\end{equation}
To go further, it is
necessary to know what is $*\tilde{\bf F}$?
Its definition  is introduced as follows. For any function $f$
to be given on
${\bf F}$-space the dual conjugation procedure is
\begin{equation}
\label{12a} *f({\bf F})=f(*{\bf F}).
\end{equation}
Then, for example,
\begin{equation}\label{zvF}
*\tilde F_{\mu\nu}({\bf F})=\tilde F_{\mu\nu}({\bf*F}).
\end{equation}
Noticing that
$$
\cos\varphi (*{\bf F})=-\cos\varphi ,\qquad \sin\varphi (*{\bf F})=
-\sin\varphi,
$$
we find by means of (\ref{zvF})
\begin{equation}\label{locdu+}
*\tilde F_{\mu \nu }=F_{\mu \nu }\sin\varphi -*F_{\mu \nu }\cos\varphi =
-e^{*\varphi }*F_{\mu \nu }.
\end{equation}

Now one can easily prove the properties
$$
(\tilde{\bf F}^2)^2+(*\tilde{\bf F}\tilde{\bf F})^2=({\bf F}^2)^2+(*{\bf FF})^2,
$$
\begin{equation}\label{cosi}
\cos\varphi=\cos\tilde\varphi ,\qquad  \sin \varphi=\sin\tilde\varphi.
\end{equation}
For instance,
\begin{eqnarray}
\cos\tilde\varphi&=&\frac{\bf\tilde F^2}{\sqrt{(\bf\tilde F^2)^2+
(*\bf\tilde F\bf\tilde F)^2}}\nonumber\\
&=&\frac{{\bf F}^2(\cos^2\varphi -\sin^2\varphi )+
2(*{\bf FF})\sin\varphi \cos\varphi
}{\sqrt{({\bf F}^2)^2+
(*{\bf FF})^2}}\nonumber\\
&=&\cos\varphi \cos2\varphi +\sin\varphi \sin2\varphi =\cos\varphi.
\nonumber
\end{eqnarray}
Taking into account (\ref{12a}), (\ref{zvF}) and (\ref{cosi}),
one can invert the formulae (\ref{locdu}),(\ref{locdu+}):
\begin{equation}\label{locdu'}
F_{\mu \nu }=e^{*\varphi }\tilde F_{\mu \nu }=e^{*\tilde\varphi }\tilde
F_{\mu \nu } , \qquad *F_{\mu \nu }= -e^{*\varphi }*\tilde F_{\mu \nu } =
-e^{*\tilde\varphi }*\tilde F_{\mu \nu }.
\end{equation}
Now the system (\ref{me}),(\ref{me'})
can be written in terms of the tensor ${\tilde F}_{\mu \nu }$
$$
\tilde F^{\mu \nu }{}{}_{,\nu }=0,\qquad
\left(e^{*\tilde \varphi }*\tilde F^{\mu \nu }\right)_{,\nu }=0,
$$
and admits, like the set (\ref{ist}),
to be written in the form
$$
\tilde F^{\mu \nu }{}{}_{,\nu }=0,\qquad *\tilde F^{\mu
\nu }{}{}_{,\nu }=\tilde\jmath^\mu,
$$
where $\jmath^\mu $
is formally  a  magnetic charge current.

In conclusion  write down the relation
$$
e^{*a\varphi}e^{*\phi}F_{\mu\nu}=e^{*(1-a)\varphi}F_{\mu\nu},
$$
where $a$
is any number.
The property to be expressed by the formula is due to the
definition (\ref{zvF})
and makes it difficult to introduce "partial" local transformation, by
means of which one could have both electrical sources and magnetic ones.

The transformations (\ref{locdu}),(\ref{locdu+}) and (\ref{locdu'})
are rather similar to discrete ones by its properties. Do they form a
group? What is deeper sense of the field $\tilde{\bf F}$
and his prototype ${\bf F}$?

\section{Duality in Non-Abelian Theory}
\subsection{The Lagrangian and Motion Equations}
The dual-invariant Lagrangian for non-Abelian fields can be chosen in the
form
$$
{\cal L}=\sqrt{({\rm Tr}{\bf
F}^2)^2+({\rm Tr}(*{\bf FF}))^2},
$$
where the fields $F_{\mu \nu }$
and the potentials $A_{\mu }$
take values in  Lie algebra of some group. The dual transformations
$$
F'_{\mu\nu}=F_{\mu\nu}\cos \theta+*F_{\mu\nu}\sin \theta=e^{*\theta}F_{\mu\nu}
,\qquad F_{\mu\nu}=e^{-*\theta}F'_{\mu\nu},
$$
$$
*F'_{\mu\nu}=*F_{\mu\nu}\cos \theta-F_{\mu\nu}\sin \theta=e^{*\theta}
*F_{\mu\nu},\qquad *F_{\mu\nu}=e^{-*\theta}*F'_{\mu\nu}
$$
are the same as in the electrodynamics. Again for convenience
 notations are introduced:
$$
a={\rm Tr}{\bf F}^2, \quad b={\rm Tr}(*{\bf FF}),\quad
 \cos\varphi=\frac{a}{\sqrt{a^2+b^2}},\quad
\sin\varphi=\frac{b}{\sqrt{a^2+b^2}},
$$
$$
\tilde F^{\mu\nu}=F^{\mu\nu}\cos
\varphi+*F^{\mu\nu}\sin\varphi=e^{*\varphi}F^{\mu\nu}.
$$
The motion equations are
$$
\tilde F^{\mu\nu} ,_\nu+i[\tilde F^{\mu\nu},A_\nu]=0.
$$
These equations are far more nonlinear than the Yang--Mills ones. To know
transformation properties of these equations one need to know how
the vector-potential $A_{\mu }$
transforms. This is the old problem~\cite{str}.
Perhaps it would be useful to apply gauge transformations.

\subsection{The Instantons}
The motion equations will be satisfied , the condition
\begin{equation}\label{insta}
\tilde F^{\mu\nu}=F^{\mu\nu}\cos \varphi+*F^{\mu\nu}\sin\varphi=0
\end{equation}
having been fulfilled.
Any solution to this equation will be referred to as a {\em generalized
instanton}. It is not difficult to see that the
Belavin-Polyakov-Schwartz-Tyupkin (BPST) instanton~\cite{bpst}
obeys  the equation. Really, the BPST instanton is the solution to the
(anti-)self-duality equation (the space-time is pseudo-Euclidean)
$$
*F_{\mu\nu}=\pm iF_{\mu\nu}.
$$
Projecting this and also the equation (\ref{insta})
onto $F_{\mu \nu }$
with tracing results in
$$
b=\pm ia,\qquad  \sqrt{a^2+b^2}=0.
$$
Since in the non-Abelian case the fields are complex  in the general
case $a,b\not=0$.
Taking into account the duality equation, the equality (\ref{insta})
can be written in the form
$$
b=\pm ia,\quad \frac{\pm i(b\mp ia)}
{\sqrt{(b+ia)(b-ia)}}F_{\mu \nu }=0.
$$
It is fulfilled for the (anti-)self-dual fields because the degree of zero
over the fraction  bar
is higher than below bar.

One can show that conventional Yang--Mills equations will be satisfied if
together with (\ref{insta})
takes place the condition
$$
*F^{\mu \nu }(\tan \varphi ),_\nu =0
$$
which is valid for the (anti-)self-dual fields.

Non-self-dual solutions were searched for in~\cite{kosh}
by means of generalization of the duality equations. It is the equation
(\ref{insta}) that may well be regarded as such a generalization.
To find new
solutions to the equation (\ref{insta}) is an important task.

\section{Dual Symmetry in Gravitation}
\subsection{Some Notations}
First one needs to introduce some notations and recall some facts writing
them in the form convenient to use further.

Curvature tensor $R_{\mu \nu \rho \sigma }$
in gravitation is  the analog of the electromagnetic strength. There are
two channels (two pair of indices) not to be quite independent  for which
the dual conjugation operation can be introduced. So $$
*R_{ijkl}=\frac{1}{2}E_{ijmn}R^{mn}{}{}_{kl},\quad
R*_{ijkl}=\frac{1}{2}R_{ij}{}{}^{mn}E_{mnkl},\quad E_{ijkl}=\frac{1}{\sqrt{-
g}}\varepsilon_{ijkl}.
$$
The properties  can be easily verified:
$$
**{\bf R}={\bf R}**=-\bf R.
$$

It is convenient to rewrite number of known properties of
the curvature tensor  in terms of the right-handed
and/or the left-handed dual-conjugate tensor.

The circular transposition identity is rewritten in the form
$$
R_{\mu \nu \rho \sigma }+R_{\mu \sigma \nu \rho }+R_{\mu \rho \sigma \nu }=0
\quad\longrightarrow \quad*R^\mu _\nu =0\mbox{ and/or }R*^\mu _\nu =0.
$$

The Bianchi identity
$$
R_{\mu \nu \rho \sigma ;\delta }+R_{\mu \nu \delta \rho ; \sigma }+
R_{\mu \nu \sigma \delta ; \rho }=0
$$
 can be written as follows
$$
*R_{\mu}{}^{\nu}{}_{\rho\sigma ; \nu }=0 \mbox{  and/or  }
R*_{\mu \nu \rho }{}{}{}^\sigma {}_{;\sigma }=0.
$$
However
$$
R*_{\mu}{}^{\nu}{}_{\rho\sigma ; \nu }\not=0 \mbox{ and/or }
*R_{\mu \nu \rho }{}{}{}^\sigma {}_{;\sigma }\not=0.
$$

Often one considers the so-called twice dual curvature
tensor (TDCT)~\cite{mtw}
$$
*R*_{\mu \nu \rho \sigma }=\frac{1}{4}E_{\mu \nu \alpha \beta }
R^{\alpha \beta \gamma \delta }E_{\gamma \delta \rho \sigma }.
$$
Using the expression for the antisymmetric $\varepsilon $-symbols product
in terms of the Kronecker $\delta $-symbols~\cite{lan},
it is not difficult to express
TDCT in terms of the curvature tensor:
\begin{equation}\label{ddtk}
*R*^{\mu \nu }{}{}_{\rho \sigma }=-R^{\mu \nu }{}{}_{\rho \sigma }+\delta^\mu_\sigma  R^\nu_\rho +\delta^\mu_\rho  R^\nu_\sigma-
\delta^\mu_\sigma R^\nu_\rho-\delta^\nu_\rho R^\mu_\sigma +\frac{1}{2}R(\delta^\nu_\rho \delta^\mu_\sigma-
\delta^\mu_\rho \delta^\nu_\sigma ).
\end{equation}

\subsection{Dual Group in Gravitation}
By means of TDCT the curvature tensor in a natural and invariant manner is
expanded into  a sum of two irreducible parts
$$
R_{\mu \nu \rho \sigma }=\cal R_{\mu \nu \rho \sigma }+
\cal S_{\mu \nu \rho \sigma },
$$
where
$$
{\cal R}_{\mu \nu \rho \sigma }=\frac{1}{2}(R_{\mu \nu \rho \sigma }
-*R*_{\mu \nu \rho \sigma }),\quad
{\cal S}_{\mu \nu \rho \sigma }=\frac{1}{2}(R_{\mu \nu \rho \sigma }
+*R*_{\mu \nu \rho \sigma }).
$$
The properties may easily be proved  to take place (sometimes  no
indices notations are used):
\begin{eqnarray}
*{\bf \cal R}={\bf \cal R}*,&&
*{\bf \cal R}*=**{\bf \cal R}={\bf\cal R}**=-{\bf\cal R},\nonumber\\
* {\bf\cal S}=- {\bf\cal S}*,&&
*{\bf\cal S}*={\bf\cal S}=-**{\bf\cal S}=-{\bf\cal S}**\nonumber.
\end{eqnarray}
For instance,
$$
*{\bf \cal R}=*\frac{1}{2}({\bf R}-*{\bf R}*)=\frac{1}{2}(*{\bf R}
+{\bf R}*)={\bf \cal R}*,
$$
$$
*{\bf \cal S}=*\frac{1}{2}({\bf R}+*{\bf R}*)=\frac{1}{2}(*{\bf R}
-{\bf R}*)=-{\bf \cal S}*
$$
etc. Since the tensors ${\cal R}$
and ${\cal S}$
transform simply in dual conjugating, the notations can be somewhat
improved. We take into account that among the left-handed and right-handed
dual-conjugate tensors (e.g., $*{\cal R},{\cal R}*$) the  independent
tensor is alone
$$
*{\bf \cal R}={\bf\cal R}*=\stackrel{*}{{\bf \cal R}},\qquad  *{\bf \cal S}=
-{\bf \cal S}*=\stackrel{*}{{\bf \cal S}}.
$$

Using the explicit expression for the TDCT, (\ref{ddtk})
one can write down the explicit expressions  for ${\cal R}$
and ${\cal S}$
\begin{eqnarray}
{\cal R}_{\mu \nu \rho \sigma }&=&R_{\mu \nu \rho \sigma }+
\frac{1}{2}\left(g_{\mu \sigma }R_{\nu \rho}+g_{\nu \rho }R_{\mu \sigma }-
g_{\mu \rho }R_{\nu \sigma }-
g_{\nu \sigma }R_{\mu \rho }\right)
\nonumber\\
&&+\frac{R}{4}(g_{\mu \rho }g_{\nu \sigma }-g_{\mu \sigma }g_{\nu \rho })
\nonumber\\
&=&C_{\mu \nu \rho \sigma }+\frac{R}{12}(g_{\mu \rho }g_{\nu \sigma }-g_{\mu
\sigma }g_{\nu \rho }),
\label{2asd}\\
{\cal S}_{\mu \nu \rho \sigma }&=&\frac{1}{2}\left(g_{\mu \rho }R_{\nu
\sigma }+ g_{\nu
\sigma }R_{\mu \rho
}-g_{\mu \sigma }R_{\nu \rho
}-g_{\nu \rho }R_{\mu \sigma }\right)\nonumber\\
&&+\frac{R}{4}(g_{\mu
\sigma }g_{\nu \rho }-g_{\mu \rho }g_{\nu \sigma }),
\nonumber
\end{eqnarray}
where $C_{\mu \nu \rho \sigma }$
is  conformal Weil's tensor.
It is also easy to obtain
\begin{equation}\label{sver}
{\cal R}_{\mu \nu }{}{}^\mu {}_{\sigma }=\frac{R}{4}g_{\nu \sigma },\qquad
{\cal R}_{\mu \nu }{}{}^{\mu \nu }=R,
\end{equation}
$$
{\cal S}_{\mu \nu }{}^\mu {}_\sigma =R_{\nu \sigma }-\frac{R}{4}g_{\nu
\sigma },\qquad  {\cal S}_{\mu \nu }{}{}^{\mu \nu }=0.
$$

Now one gets ready to introduce a dual group in gravitation. Let us
consider the
transformation
\begin{equation}\label{grou}
{\bf R}\to {\bf
R'}:\quad{\bf R'}=e^{*\alpha}{\bf R}e^{*\beta},
\end{equation}
where $\alpha $
and $\beta $
are independent dual angles. More precisely, the somewhat symbolic notation
(\ref{grou}) means
\begin{equation}\label{grou'} {\bf R'}={\bf
R}\cos\alpha\cos\beta+*{\bf R}\sin\alpha\cos\beta+{\bf
R}*\cos\alpha\sin\beta+*{\bf R}*\sin\alpha\sin\beta.
\end{equation}
Quite clear, the transformations of kind (\ref{grou'})
form an Abelian two-parameter group.

Next, one finds that the tensors ${\cal R,\cal S}$
transform {\em simply} when acted by the group (\ref{grou'})
\begin{equation}
\label{irr} {\bf\cal R'}=e^{*(\alpha +\beta
)}{\bf\cal R}={\bf\cal R}e^{*(\alpha +\beta )},\quad {\bf\cal
S'}=e^{*(\alpha -\beta )}{\bf\cal S}={\bf\cal S}e^{-*(\alpha -B)}
\end{equation}
and actually realize an irreducible representation of the dual group in
gravitation.

By analogy with the dual electrodynamics, as a consequence of (\ref{irr})
we immediately obtain  two dual gravitational invariants:
$$
I_1=({\bf
\cal R}^2)^2+(\stackrel{*}{{\bf\cal R}}
\stackrel{\phantom{*}}{\bf\cal R})^2,\quad I_2=
({\bf\cal S}^2)^2+(\stackrel{*}{\bf\cal S}
\stackrel{\phantom{*}}{\bf\cal S})^2.
$$
Here ${\cal R}^2={\cal R^{\mu \nu \rho \sigma }}{\cal R}_{\mu \nu \rho
\sigma }$
etc. Perhaps, it would be far more convenient to use the linear
combinations of these invariants which are
expressed in terms of this curvature tensor:
\begin{eqnarray}
J_1&=&({\bf R}^2)^2+(*{\bf RR})^2+({\bf RR}*)^2+(*{\bf RR}*)^2,\nonumber\\
J_2&=&{\bf R}^2(*{\bf R}{\bf R}*)-
(*{\bf R}{\bf R})({\bf R}{\bf R}*),\nonumber
\end{eqnarray}
where in reality
$$
*{\bf R}{\bf R}={\bf R}{\bf R}*,
\quad *{\bf R}{\bf R}*={\bf R}(*{\bf R}*).
$$

\subsection{The Gravitational Instantons}
\subsubsection{Discussion}
The current state of the gravitational instantons question seems to be
somewhat
intricate. Penrose's~\cite{pen} instantons (nonlinear gravitons) are
the (anti)self-dual complex solutions to  Einstein's equations. Hawking
introduces the instantons as Euclidean solutions to  Einstein's equations with
finite action~\cite{haw}.  This matter  is reviewed e.g. in~\cite{gib}.
In either event gravitational instantons are related to solutions to
Einstein's
equations in the {\em Euclidean} space
$$
R_{\mu \nu }=\lambda g_{\mu \nu }
$$
and should obey the (anti-)self-duality equations to be understood as
follows
$$
*C_{\mu \nu \rho \sigma }=\pm C_{\mu \nu \rho \sigma },
$$
where $C_{\mu \nu \rho \sigma }$ is
Weil's tensor. \footnote{This point has been cleared up to me by A.
Popov.} As seen, for example, from non-Abelian theory, the instantons, to a
certain degree, do not depend on dynamics. They rather display deeper
kinematic-topological properties. For  Einstein's gravitation it is not
the point.

It is not worthwhile  to relate  instantons to any dynamical equations,
  to  Einstein's ones in particular. These equations, as
distinct
from  Yang--Mills ones, place too hard restrictions on the curvature
tensor  from  the point of view of the (anti-)self-duality properties.

Let
 us call as  gravitational  instantons  the solutions to the duality
equations in {\em pseudo-Euclidean} space for the tensors ${\cal R}$
and {\cal S}
\begin{equation}\label{autodu}
\stackrel{*}{\bf\cal R}=\pm i{\bf\cal R},
\qquad \stackrel{*}{\bf\cal S}=\pm i{\bf\cal S}.
\end{equation}
These equations are quite equivalent to the duality conditions in the
Yang--Mills theory. Of course, the
${\bf R}$-space is real    and the equations (\ref{autodu})
are reduced to
\begin{equation}\label{kon}
{\cal R}_{\mu \nu \rho \sigma }=0,
\qquad {\cal S}_{\mu \nu \rho \sigma }=0.
\end{equation}
So, the real gravitational instantons are determined by the equations
(\ref{kon}). To avoid misunderstanding it should be emphasized that the
equations (\ref{kon})  must not be regarded as a system (set).

Written in another form, such equations are given in the book~\cite{kon}
and they were obtained otherwise.
Solutions to these equations (which are not  obtained and are not presented
in the book) are referred to as twice (anti-)self-dual ones, which are
similar
to usual instantons by their properties.

It follows from our approach that they are usual gravitational
instantons.

Next a few central-symmetry solutions are shown.

\subsubsection{The 4-central-symmetric solutions}
\paragraph{The Metric Choice.}
Let us look for the solutions to (\ref{kon})
as a metric of the form
$$
ds^2=e^{\nu(\rho)}d\rho^2-\rho^2[d\psi^2+\sinh^2\psi(d\theta^2+
\sin^2\theta d\phi^2)],
$$
where
$$
ds_0^2=d\rho^2-\rho^2[d\psi^2+\sinh^2\psi(d\theta^2+\sin^2\theta d\phi^2)]
$$
is a 4-spherical flat metric. Calculating the curvature tensor  gives
four nonzero (diagonal) components of the tensor $R_{\mu \nu }$
and six nonzero
(diagonal) components of the tensors $R_{\mu \nu \rho \sigma }$.

\paragraph{$\protect\bf{{\cal R}_{\bf\mu \nu \rho \sigma }=0.}$}
Six components of the equation that do not turn into identities   reduce
to the only one of the {\em first} order
$$\nu '(\rho )=\frac{2}{\rho }(1-e^\nu )
$$
The equation is easily solved:
$$
e^{\nu (\rho )}=\frac{\rho ^2}{\rho ^2-C}.
$$
The metric is getting flat if $\rho \to\infty$
or $C=0$.

\paragraph{$\protect\bf{{\cal S}_{\mu \nu \rho \sigma }=0.}$}
Six components of the equation   not to be reduced to identities reduce to
a differential {\em first} order equation  alone
$$
\nu '(\rho )=\frac{2}{\rho }(e^\nu-1),
$$
which has the solution
$$
e^\nu =\frac{1}{1-C\rho ^2}.
$$
If $\rho \to0$
or $C=0$
the metric gets flat.

\paragraph{$\protect\bf{R_{\mu \nu }=0.}$}
Einstein's equations in empty space have a trivial solution only: $e^\nu
=0$.
This  merely emphasizes the fact that the gravitational instantons are
poorly compatible with Einstein's equations.
The solutions described in this section are quite analogous to the
spherical-symmetric BPST's instanton~\cite{bpst}.

\subsubsection{The Static Central-Symmetric Solutions}
\paragraph{The Metric Choice.}
We search for a solution as follows (the metric is just as in Landau and
Lifshits~\cite{lan}):
$$
ds^2=e^{\nu(r)}dt^2-e^{\lambda(r)}dr^2-r^2(d\theta^2+\sin^2\theta d\phi^2).
$$
As a result of the curvature tensor calculation it turned out
to consist of four nonzero
(diagonal) components of the tensor $R_{\mu \nu }$ and six nonzero
(diagonal) ones of $R_{\mu \nu \rho \sigma }$.

\paragraph{$\protect\bf{{\cal R}_{\bf\mu \nu \rho \sigma }=0.}$}
Six equations not to be reduced to identities reduce to the only
differential equation of the {\em second} order
$$
\lambda =\nu ,\qquad \nu ''(r)=\frac{2}{r^2}(e^\nu  -1).
$$
This is a rather nontrivial equation. It can be exactly solved~\cite{kam}.
The solution is represented in two forms:
$$
e^\nu
=\frac{C_1^2}{2}r^2\sin^{-2}\left[\frac{C_1}{\sqrt{2}}(r-C_2)\right]
$$
or
$$
e^\nu
=\frac{C_1^2}{2}r^2\sinh^{-2}\left[\frac{C_1}{\sqrt{2}}(r-C_2)\right].
$$
With the constant $C$
equal to zero, the solution becomes more simple:
$$
e^\nu =\frac{r^2}{(r-C)^2}.
$$
This solution is asymptotically flat at $r\to\infty$
and for a small $r$:
$$
e^\nu \sim r^2,\qquad r\to 0.
$$

\paragraph{$\protect\bf{{\cal S}_{\mu \nu \rho \sigma }=0.}$}
Six equations not to be identities reduce to the only equation of the {\em
second} order
$$
\lambda =-\nu ,\qquad \nu ''(r)+\nu '^2(r)=\frac{2}{r^2}(1-e^{-\nu }).
$$
which is solved simply:
$$
e^\nu =1+C_1r^2+\frac{C_2}{r}.
$$
It is seen that this solution contains Schwartzschild's solution
(if $C_1=0$).
For large $r$
the first metric coefficient goes to infinity, what  points out that
the metric may  be closed.

\subsubsection{Generalized Instantons}
By analogy with the dual electrodynamics and non-Abelian theory, the
generalized instantons are defined as follows (it is not a system
of equations):
\begin{equation}\label{obins}
{\cal R}_{\mu \nu \rho \sigma }\cos \varphi +\stackrel{*}{\cal R}_{\mu
\nu \rho \sigma }\sin \varphi =0,
\end{equation}
\begin{equation}\label{obins'}
{\cal S}_{\mu \nu \rho \sigma }\cos \psi +\stackrel{*}{\cal S}_{\mu
\nu \rho \sigma }\sin\psi=0,
\end{equation}
where
\begin{eqnarray}
\cos\varphi=\frac{a}{\sqrt{a^2+b^2}},&&\sin\varphi=\frac{b}{\sqrt{a^2+b^2}},
\nonumber\\
\cos\psi=\frac{c}{\sqrt{c^2+d^2}},&&\sin\psi=\frac{d}{\sqrt{c^2+d^2}},
\nonumber
\end{eqnarray}
$$
a={\bf\cal R}^2,\quad
b=\stackrel{\phantom{*}}{\bf\cal R}\stackrel{*}{\bf\cal R},\quad
c={\bf\cal S}^2,\quad
d=\stackrel{\phantom{*}}{\bf\cal S}\stackrel{*}{\bf\cal S}.
$$
It follows from (\ref{obins}),(\ref{obins'})
$$
a^2+b^2=0,\qquad c^2+d^2=0
$$
or $a=0,\,b=0,\,c=0,\,d=0$.
The instantons ${\cal R}=0$
and ${\cal S}=0$
are easily seen to obey the equations (\ref{obins}),(\ref{obins'}).
Perhaps, there exists some way to complexify
${\bf R}$-space, so that the instanton notion in gravitation would be more
comprehensive as it should be in non-Abelian theory.

\subsection{The Gravitation Equations}
If there are tensors that possess the basic symmetries of the curvature
tensor, metric and the energy-momentum tensor of matter, new gravitational
equations could be constructed. Let us begin from the trivial but visual
example. It is possible to construct tensor by means of the metric and
the energy-momentum 
tensor which has the curvature tensor symmetries.
The following equation is postulated:
\begin{equation}\label{triv}
R_{\mu \nu \rho \sigma }=const (g_{\mu \rho }T_{\nu \sigma }+g_{\nu \sigma }
T_{\mu \rho }-g_{\mu \sigma }T_{\nu \rho }-g_{\nu \rho }T_{\mu \sigma }).
\end{equation}
Is this equation good or bad?
It is bad as it follows from below. Let the energy-momentum tensor be
concentrated at finite range of the space. Out of the range the equation
is given by
$$
R_{\mu \nu \rho \sigma }=0.
$$
Thus, the equation (\ref{triv})
predicts the absence of gravity wherever the matter is absent.

Let us try once more to find the gravitational equation using the twice
anti-self-dual part of the curvature tensor
\begin{equation}\label{nord}
{\cal R}_{\mu \nu \rho \sigma }=A(g_{\mu \rho }T_{\nu \sigma }+g_{\nu \sigma }
T_{\mu \rho }-g_{\mu \sigma }T_{\nu \rho }-g_{\nu \rho }T_{\mu \sigma }).
\end{equation}
One can show that in the long run the equation reduces
to the conformal flat
Nordstr$\ddot{\mbox{o}}$m's theory~\cite{nor}.
Really, contracting  it
with respect to the indices $\mu $
and $\rho $
and taking into account (\ref{sver}),
we obtain
\begin{equation}\label{1}
\frac{1}{4}Rg_{\nu \sigma  }=A(2T_{\nu \sigma }+g_{\nu \sigma }T).
\end{equation}
One more contraction gives
\begin{equation}\label{2}
R=6AT.
\end{equation}
Expressing $T_{\nu \sigma }$
from (\ref{1})
and substituting it to (\ref{nord}),
also taking into account (\ref{2}),(\ref{2asd}), we find eventually
$$
{\cal R}_{\mu \nu
\rho \sigma }=C_{\mu \nu \rho \sigma }+ \frac{R}{12}(g_{\mu \rho }g_{\nu
\sigma }-g_{\mu \sigma }g_{\nu \rho })= \frac{R}{12}(g_{\mu \rho }g_{\nu
\sigma }-g_{\mu \sigma }g_{\nu \rho }).
$$
Thus, $C_{\mu \nu \rho \sigma }=0$.
Together with (\ref{2})
this equation is a formulation of  Nordstr$\ddot{\mbox o}$m's
conformal flat theory~\cite{mtw}
which, for example, predicts no deviation of light in the gravity field.

Being consequent, we have to try constructing an equation by means of the
tensor ${\cal S}$:
$$
{*\cal S}_{\mu \nu \rho \sigma }=
B(g_{\mu \rho }T_{\nu \sigma }+g_{\nu \sigma }
T_{\mu \rho }-g_{\mu \sigma }T_{\nu \rho }-g_{\nu \rho }T_{\mu \sigma }).
$$
One can show that this equation reduces to
$$
R_{\mu \nu }-\frac{R}{4}g_{\mu \nu }=2BT_{\mu \nu },\qquad T=0.
$$
It would be of interest to consider the system: electromagnetic field --
gravitation starting from these equations rather than the Maxwell-Einstein
ones.

\subsection{Duality and Variational Principle in Gravitation}
We  concerned the dynamical aspects just in the previous subsection
trying to construct gravitational equations by means of the dual-symmetric
parts of the curvature tensor only. It is of interest to establish a
variational principle which is compatible with duality, say, in the way to be
similar to the dual electrodynamics variational principle. As has been
noted, Einstein's equation,  and consequently,
Hilbert's variational principle
are poor compatible with duality.

However, direct attempt to create a gravitational theory with the dual
symmetry to be involved faces troubles.

Let us treat maintaining the analogy to electrodynamics. It is known that
the quadratic Lagrangian electrodynamics equations are dual invariant. In
the gravitational case that all would have been analogous if  the equations
\begin{equation}\label{eqgr}
R_{\mu \nu \rho }{}{}{}^\sigma {}_{;\sigma }=0\qquad,
R*_{\mu \nu \rho }{}{}{}^\sigma {}_{;\sigma }\equiv0
\end{equation}
had taken place.
But for the quadratic in ${\bf R}$
Lagrangian one obtains the equation
\begin{equation}\label{eqgr'}
R_{\mu (\nu \rho) }{}{}{}^\sigma{}_{;\sigma }=0
\end{equation}
rather than the first one of (\ref{eqgr})
Christoffel's symbols are suggested to be related to metric in the usual
way but connections are varied rather than the metric.
However, the left-hand side of (\ref{eqgr'})
is identically zero on account of the circular transposition identity. Thus
we have no variational principle leading to the equations (\ref{eqgr})
and at this point the analogy to electrodynamics already vanishes.

Perhaps, one should consider {\em nonsymmetric} connections introducing in
this way torsion. In any case this would enable one to avoid the equation
(\ref{eqgr'}).
None forbid however to compound the dual-symmetric objects by means of the
tensors ${\cal R}$
and ${\cal S}$.
In effect this has been done in the subsection about the gravitational
instantons. It is possible also to use directly of two gravitational dual
invariants. We have to repeat, however, that besides the problem to choose
the Lagrangian, there is another difficulty. If the basic dynamical
variables are symmetric connections  related to the metric
in the usual fashion
then the dual invariant Lagrangian theory will be ugly in view of
(\ref{eqgr'}).

\section{Conclusion}

In the paper an attempt to consider the consequences of the theory with the
incorporated dual symmetry has been made. The dual symmetry in gravitation
has been investigated as well. Such an approach to gravitation has not
appeared  before. The dual electrodynamics is the elegant
nonlinear model which might be related to the linear Maxwell
electrodynamics. The local dual angle arises in a natural fashion in this
theory. This angle enables introducing the local dual transformations. By
means of these transformations the theory may be reformulated in terms of
the magnetic charge.   The equations have been considered the solutions to
which were referred to as the generalized instantons (for the
BPST-instanton obeys these equations). Nontrivial gravitational instantons
have been found.  Irreducible representations of the dual group in
gravitation give new opportunities to establish  new gravitational
equations. The problem related to variational principle
notion for the dual-symmetric gravitation has been discussed.

The tasks of interest should be noted.
\begin{itemize}
\item
Searching for solutions to the dual electrodynamics and non-Abelian theory
(the instantons as well). The problem to find the {\em asymptotically}
central-symmetric solutions is of great importance.
\item
The transformation property in the non-Abelian theory (in particular for
vector-potential) under the dual group.
\item
Establishing the dual-symmetric variational principle in gravitation.
Perhaps, the torsion should be included in the theory  to obtain
self-consistent theory.
\end{itemize}

The author is thankful to the organizers and participants
of the seminar "Quantum
Symmetries, Gravitation and Strings" of BLTP JINR, namely
to B.Barbashov, A.Pestov,
A.Popov, V.Dubovic and others, for the opportunity
to give a talk and useful
discussions.

\end{document}